\documentclass{article}

\usepackage{arxiv}

\usepackage[utf8]{inputenc} 
\usepackage[T1]{fontenc}    
\usepackage{hyperref}       
\usepackage{url}            
\usepackage{booktabs}       
\usepackage{amsfonts}       
\usepackage{nicefrac}       
\usepackage{microtype}      
\usepackage{lipsum}		
\usepackage{graphicx}
\usepackage{natbib}
\usepackage{doi}

\usepackage{framed,multirow}

\usepackage{amssymb}
\usepackage{latexsym}
\usepackage{lineno}
\usepackage{graphicx}
\usepackage{amsmath,amstext,amssymb,color,amsthm}
\usepackage{float}
\usepackage{mwe}
\usepackage{comment}
\usepackage{booktabs}
\usepackage{textcomp}
\usepackage{algorithm,algorithmic}
\usepackage{array}
\usepackage{url}
\usepackage{caption}
\usepackage{subcaption}
\usepackage{multirow}
\usepackage{comment}
\usepackage{float}
\usepackage{soul}

\usepackage{tikz}
\usetikzlibrary{positioning}
\usetikzlibrary{3d}
\usetikzlibrary{matrix}
\usetikzlibrary{decorations.text}
\usetikzlibrary{spy}

\makeatletter
\tikzoption{canvas is plane}[]{\@setOxy#1}
\def\@setOxy O(#1,#2,#3)x(#4,#5,#6)y(#7,#8,#9)%
{\def\tikz@plane@origin{\pgfpointxyz{#1}{#2}{#3}}%
	\def\tikz@plane@x{\pgfpointxyz{#4+#1}{#5+#2}{#6+#3}}%
	\def\tikz@plane@y{\pgfpointxyz{#7+#1}{#8+#2}{#9+#3}}%
	\tikz@canvas@is@plane
}
\makeatother
\tikzoption{canvas is xy plane at z}[]{%
	\def\tikz@plane@origin{\pgfpointxyz{0}{0}{#1}}%
	\def\tikz@plane@x{\pgfpointxyz{1}{0}{#1}}%
	\def\tikz@plane@y{\pgfpointxyz{0}{1}{#1}}%
	\tikz@canvas@is@plane
}
\makeatother

\usepackage{url}
\usepackage{xcolor}
\usepackage{hyperref}
\definecolor{newcolor}{rgb}{.8,.349,.1}

\title{Recursive Deep Prior Video: a Super Resolution algorithm for Time-Lapse Microscopy of organ-on-chip experiments}

\date{} 					

\author{ \href{https://orcid.org/0000-0002-1475-2751}{\includegraphics[scale=0.06]{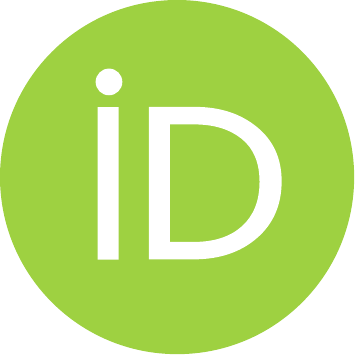}\hspace{1mm}Pasquale Cascarano}\thanks{\textit{email:} pasquale.cascarano2@unibo.it} \\
	Department of Mathematics \\ 
	University of Bologna \\ 
	Piazza di Porta S. Donato 5, \\
	Bologna, Italy, 40126

	\And
	
	\hspace{1mm} Maria Colomba Comes \thanks{\textit{email:} maria.colomba.comes@uniroma2.it}  \\
	Department of Electronic Engineering \\ 
	University of Tor Vergata \\ 
	Via del Politecnico 1, \\
    Rome, Italy,00133
	
	\And
	
	\hspace{1mm} Arianna Mencattini \\
	Department of Electronic Engineering \\ 
	University of Tor Vergata \\ 
	Via del Politecnico 1, \\
    Rome, Italy,00133
    
    \And
	
	\hspace{1mm} Maria Carla Parrini \\
	Institute Curie Centre de Recherche \\ 
	Paris Sciences et Lettres Research University \\ 
    Paris, France, 75005

        \And
    
    \hspace{1mm} Elena Loli Piccolomini \\
	Department of Computer Science and Engineering \\ 
	University of Bologna \\ 
	Mura Anteo Zamboni 7, \\
    Bologna, Italy,40126
    
     \And
      
    \hspace{1mm} Eugenio Martinelli \\
	Department of Electronic Engineering \\ 
	University of Tor Vergata \\ 
	Via del Politecnico 1, \\
    Rome, Italy,00133

}

\renewcommand{\shorttitle}{RDPV: a Super Resolution algorithm for TLM}

\begin{document}
\maketitle

\begin{abstract}
	Biological experiments based on organ-on-chips (OOCs) exploit light Time-Lapse Microscopy (TLM) for a direct observation of cell movement that is an observable signature of underlying biological processes. A high spatial resolution is essential to capture cell dynamics and interactions from recorded experiments by TLM. Unfortunately, due to physical and cost limitations, acquiring high resolution videos is not always possible. To overcome the problem, we present here a new deep learning-based algorithm that extends the well known Deep Image Prior (DIP) to TLM Video Super Resolution (SR) without requiring any training. The proposed Recursive Deep Prior  Video (RDPV) method introduces some novelties. The weights of the DIP network architecture are initialized for each of the frames according to a new recursive updating rule combined with an efficient early stopping criterion. Moreover, the DIP loss function is penalized by two different Total Variation (TV) based terms. The method has been validated on synthetic, i.e., artificially generated, as well as real videos from OOC experiments related to tumor-immune interaction. Achieved results are compared with several state-of-the-art trained deep learning SR algorithms showing outstanding performances.
\end{abstract}


\section{Introduction}\label{sec:Intro}
\begin{figure*}[!t]
\centering
\includegraphics[scale=.4]{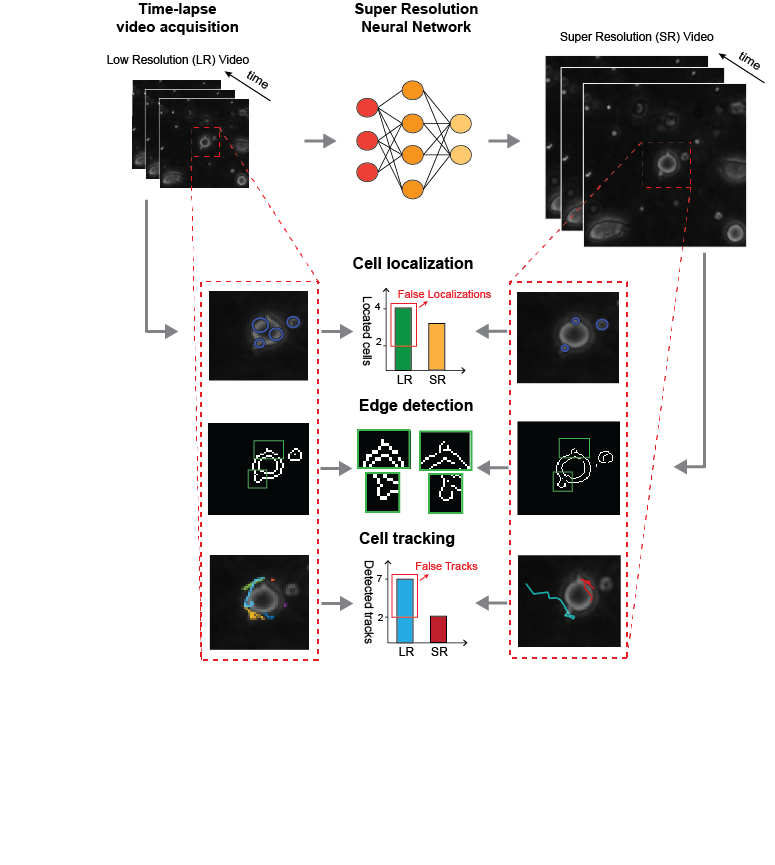}
\caption{\textbf{More effective cell motility evaluation moving from Lower Resolution to Super Resolution videos.} The number of false occurrences during the cell localization phase is reduced (Cell localization); cells in contact are separated thanks to an efficient edge detection (Edge detector); cell trajectories are effectively constructed (Cell tracking).}
\label{fig:fig_1}
\end{figure*}

Light Time-Lapse Microscopy (TLM) imaging is successfully used to acquire and record biological experiments based on Organ-On-Chip (OOC) platforms which are miniaturized microfluidic devices mimicking in-vitro complex 3D cellular micro-environments \citep{polini2014organs}, such as cell migration \citep{kabla2012collective} or multi-cellular interaction \citep{businaro2013cross,agliari2014cancer}. 
After acquisition by TLM, the live cell videos are analyzed by means of sophisticated computerized algorithms with the aim of finding biological insights embedded within cell motility. The way in which cell move, indeed, has been discovered meaningful to understand biological processes, such as wound healing \citep{shaw2009wound} and morphogenesis \citep{friedl2009collective} but also cancer growth and spread of metastasis \citep{friedl2003tumour}. Usually, automated exploitation of such devices includes localizing and tracking cells through increasingly cutting-edge single particle tracking software \citep{biselli2017organs,nguyen2018dissecting, payer2019segmenting,arbelle2018probabilistic}. Cell trajectories are then coded in time and space domain by extracting some motility descriptors useful to uncover and quantitatively evaluate the response to target therapeutic agents \citep{di2019learning, parlato20173d}.
However, the reached biological conclusions can be compromised by the experimental set-up of TLM in terms of spatiotemporal resolution \citep{Comes2019,beltman2009analysing}. First of all, a suitably high frame rate is required to entirely capture and then to accurately estimate the duration of biological events such as apoptosis or cell-cell interaction. Not less relevant, a high spatial resolution implies many benefits for the video analysis. Spatial resolution of a light TLM video frame refers to the number of pixels composing the image for a fixed field of view. The higher pixel density, characterizing the so called High Resolution (HR) images with respect to Low Resolution (LR) ones, reveals to be essential to better distinguish particles within a video and hence to effectively dissect intricate biological phenomena involving multiple cell populations. Multi-cellular interaction may constitute an example: during interaction, the distance among cells reduce until they overlap. A high spatial resolution may decrease cell detection and tracking errors showing more defined and detailed shapes and edges. As a result, a more reliable tracking positively affects the trustworthiness of the motility descriptors extracted from the trajectories thus uncovering unbiased biological findings.
Unfortunately, acquiring HR images is not always possible, due to the high cost of the high performing acquisition instruments and physical constraints, such as the optical zoom provided by the camera. However, if a higher magnification is feasible, it does not certainly guarantee a better image clarity. 
In addition, HR images related to long-term experiments, i.e., from hours to days, can reach a size from tens to hundreds of gigabytes, thus requiring very high processing capabilities. 
A fair compromise between high computational requirements and the preservation of the biological informative content may be represented by the adoption of Super Resolution (SR) algorithms: experiments can be acquired at a LR and then post-processed  by means of SR algorithms. In the classical LR image formation model, the LR data is assumed to be a down-sampled and noisy version of the HR data. By passing from HR to LR image, some high-frequency details are lost, and, as consequence, the inverse process of reconstruction from LR to HR image represents an ill-posed problem, namely, some HR images result to be related to a single LR image. To effectively reconstruct the HR image from the LR one, many of the most efficient SR methods proposed in the literature require a-priori information on the HR data that should be retrieved \citep{yue2016}. Other SR methods utilize deep learning to learn the statistical correlation between LR images and their HR counterparts from an external database of training examples \citep{EDSR,RCAN,proSR,ESRGAN}. In the last decades, SR algorithms have been covered a wide spectrum of applications, from medical imaging \citep{ravi2019, rueda2013,isaac2015}, thermal imaging \citep{cascarano2020,rivadeneira2019}, remote sensing imaging \citep{yang2015}, to security imaging \citep{lin2005}. To the best of our knowledge, SR algorithms have been also formulated for fluorescence microscopy applications \citep{leung2011review, ravi2019,godin2014super,wang2019deep,weigert2018content}, whereas they have been not developed for light microscopy applications. 
In this paper, we propose a novel super resolution framework based on the recently developed Deep Image Prior (DIP) \citep{Ulyanov} and applied to light TLM videos representing living cell migration and interaction. We introduce three algorithms: the Deep Prior Video (DPV), which is an extension of DIP to video frames, the Recursive Deep Prior Video (RDPV), based on a recursive updating rule for the weights of the network, and, finally, a regularized version of RDPV based on Total Variation.
As well as for the DIP framework, the proposed approaches exploit the prior knowledge on the solution carried by the Convolutional Neural Network (CNN) structures. One of the most striking novelty of the proposed methods is the fact that it does not require any training step, being totally unsupervised, thus drastically reducing the time of running and accelerating the experimental response on cell behaviors from TLM experiments. Three additional main novelties are introduced. Firstly, a sharing parameters technique among consecutive video frames is implemented: RDPV takes in input one frame of a TLM video at a time and uses the knowledge of previous super resolved video frames to reconstruct the new one through a new recursive updating rule for the weights of the network. Secondly, a stopping rule is fixed to reduce the running time and to avoid overfitting, which is typical in the classical DIP framework.  Finally, inspired by \citep{liu2019image}, we boost the performances of the RDPV by adding two commonly used Total Variation (TV) terms, namely isotropic TV (RDPV-TVi) and anisotropic TV (RPV-TVa) \citep{Rudin1992}. 
In Fig.~\ref{fig:fig_1}  an example of video frame reconstruction with RDPV is shown. The favorable effect of super resolution on cell localization, edge map detection and cell tracking are also highlighted. The application of the RDPV algorithm allows us to reduce the false occurrences during the cell localization phase (Cell localization in Fig.~\ref{fig:fig_1}), to successfully separate cells in contact (Edge detector in Fig.~\ref{fig:fig_1}) and to effectively construct cell trajectories (Cell tracking in Fig.~\ref{fig:fig_1} ). 
The proposed DPV, RDPV and its TV-based variants are validated on synthetic videos, i.e., artificially generated, and real videos from organ-on-chip experiments. As results, RDPV and its regularized versions greatly improve the performances obtained by the application of the classical DIP approach on videos, DPV. The new methods are also compared with the state-of-the-art of trained deep learning-based SR algorithms achieving very promising results \citep{EDSR,RCAN,proSR,ESRGAN}. The evaluation of the SR image quality improvement is performed by using classical metrics such as Peak-Signal-to-Noise-Ratio (PSNR) and the Structural SIMilarity index (SSIM) \citep{hore2010}. In addition, the reliability of the biological information supplied by LR and the corresponding super resolved videos are assessed.

\section{A Brief Survey on Super Resolution} \label{related works}

In the field of SR, the Single Image Super Resolution (SISR) task, which is the problem of reconstructing an HR image from a single LR image, has mostly caught the attention among the researchers. In the discrete setting, the single image SR observation model can be fixed as follows: 

\begin{equation}\label{eq:SISR_problem}
    S x + \eta = y, 
\end{equation}

where $S \in \mathbb{R}^{NM \times L^{2}NM}$ is the degradation matrix which links the HR image $x \in \mathbb{R}^{L^{2}MN \times 1}$ to the LR image $y \in \mathbb{R}^{MN \times 1}$, both reshaped as vectors. The positive integer $L$, also known as magnification factor, is the square root of ratio between the dimensions of the HR and LR images. 
Moreover, $\eta \in \mathbb{R}^{MN \times 1}$ is the Additive White Gaussian Noise (AWGN) component of standard deviation $\sigma$, which corrupts the LR measurement. Mathematically, the single SR problem is known to be strongly ill-posed \citep{yue2016}, therefore the solution results to be sensitive to the presence of noise in $y$. Moreover, the linear system in \eqref{eq:SISR_problem} is an underdetermined system, i.e., many HR images are related to the acquired LR image $y$; hence inverting the process described by \eqref{eq:SISR_problem} has not a unique solution. In the literature, many approaches have been proposed in order to get a good estimate of $x$. Among them, the interpolation-based methods reallocate the LR image pixels on a finer grid, then the missing pixel values are estimated using interpolation techniques such as Bicubic interpolation and Lanczos resampling \citep{duchon1979}. Even if the interpolation-based methods are computationally cheap, they suffer from accuracy issues since they are not capable to retrieve the missing high frequency details. 
Another common approach in the literature is to add some prior information on the desired HR solution $x$, in order to shrink the space of the possible solutions. This is done in different ways by the algorithms belonging to the class of reconstruction-based SISR and learning-based SISR methods which are currently the state-of-the-art.
The reconstruction-based methods rewrite the problem \eqref{eq:SISR_problem} as the following unconstrained minimization problem: 
\begin{equation} \label{eq:rec_based}
    x^{*}=\arg \min_{x} \lVert y - Sx \rVert_{2}^{2} + \lambda \rho(x).
\end{equation}
The objective function in \eqref{eq:rec_based} encodes both the information on the fixed observation model  (type of noise, blur, downsampling, etc.) and the prior information on the solution (smoothess, sparsity, etc.) by the first and the second term in the sum, respectively. The scalar parameter $\lambda$ is an hyperparameter which represents a trade-off between the two terms \citep{yue2016}.
Despite their good performances, their flexibility and their strong mathematical underpinnings, the quality of the reconstruction computed by those methods degrades rapidly when the magnification factor $L$ increases.
Learning-based SISR methods utilize Deep Neural Network (DNN) architectures to learn the statistical correlation between the LR and the HR image in a supervised framework given a training set of examples. They attempt to solve the following minimization problem: 
\begin{equation}\label{eq:DL}
    \theta^{*} \in \arg \min_{\theta} \mathcal{L}(f_{\theta}(y), x) \quad \text{for} \ y \in Y, x \in X, 
\end{equation}
where $f_{\theta}$ is a fixed DNN architecture with weights $\theta$, $\mathcal{L}$ is a  loss function and  $\lbrace (y,x) \rbrace_{y \in Y, x \in X}$ is a large given set of LR and HR  example pairs.  Then, for a generic LR image $\Tilde{y}$ the HR reconstruction $\Tilde{x}$ is obtained as $\Tilde{x}=f_{\theta^{*}}(\Tilde{y})$.
The success of supervised learning-based methods is due to the capacity of these models to handle a big amount of data and to capture image statistics. 
Among the more powerful deep learning methods for SISR we mention the Enhanced Deep Super Resolution (EDSR) neural network \citep{EDSR}, and the Residual Channel Attention Network (RCAN) \citep{RCAN}, both based on CNN structures, and the Enhanced Super Resolution Generative Adversarial Networks (ESRGAN) \citep{ESRGAN} based on adversarial training. However, their outstanding performances require a varied set of examples and consequently an overstated computational effort. Moreover, in most cases the trained deep architectures are specialized for a fixed magnification factor $L$, requiring to repeat the training if a different magnification factor is needed. It is worth saying that is quite difficult in real applications either to acquire images of the same scene with different resolutions or dispose of a varied data set. In addition, they can exhibit instability issues due to the presence of noise in the LR data \citep{wang2020}. \\
SR techniques have also been developed to enhance the spatial resolution of the videos by reconstructing an HR version of each LR frame constituting the video. Most of the real-world Video Super Resolution (VSR) approaches require a very high-temporal acquisition frequency, i.e., from 15 to 60 frames per second  \citep{wang2020,kappeler2016,Nah_2019}. Since in this work we focus on TLM videos acquired at fixed time step, from seconds \citep{Comes2019} to minutes \citep{nguyen2018dissecting}, we only refer to SISR algorithms as competing methods, neglecting the VSR approaches.

\section{Method}\label{method}

\subsection{Deep Image Prior}

In the field of SR, the difficulty of constructing a set of LR-HR image pairs has prompted researchers to inspect unsupervised learning-based approaches. One of the most interesting works in this direction is Deep Image Prior (DIP) \citep{Ulyanov}. In their pioneering work, the authors prove that a CNN structure is capable to capture the image features without employing any training set. Indeed, it has been shown that natural images are more easily reproducible than random noise by a deep CNN.  Fixed a CNN structure $f_{\theta}$, with weights $\theta$, and a random vector $z \in \mathbb{R}^{sM \times sN}$, DIP attempts to reconstruct the HR image $x^{*}$ by solving the following optimization problem:

\begin{equation} \label{DIP_minimization}
    \theta^{*} \in \arg \min \lVert y - S f_{\theta}(z) \rVert^{2}_{2}, \quad x^{*}=f_{\theta^{*}}(z).
\end{equation}

The CNN $f_{\theta}$ is randomly initialized and then the weights are iteratively optimized so that the output of the network is as close to the unknown HR target as possible. It is well known that the choice of the iteration number is crucial to not reproduce the noise affecting the initial LR data y in the solution $x^{*}$ \citep{Ulyanov}. From a theoretical point of view, Ulyanov et al. have shown that, in the first iterates of the optimization process, the architecture of a CNN reproduces HR noise-free images. So far, DIP has shown comparable performances with respect to other supervised learning-based methods in the field of SISR; however it has some drawbacks that must be solved, e.g., a suitable stopping rule for the iterative process is needed and the choice of a proper CNN architecture must be accounted.

\subsection{Proposed methods}

\begin{figure}
    \centering
	\includegraphics[width=0.8\textwidth]{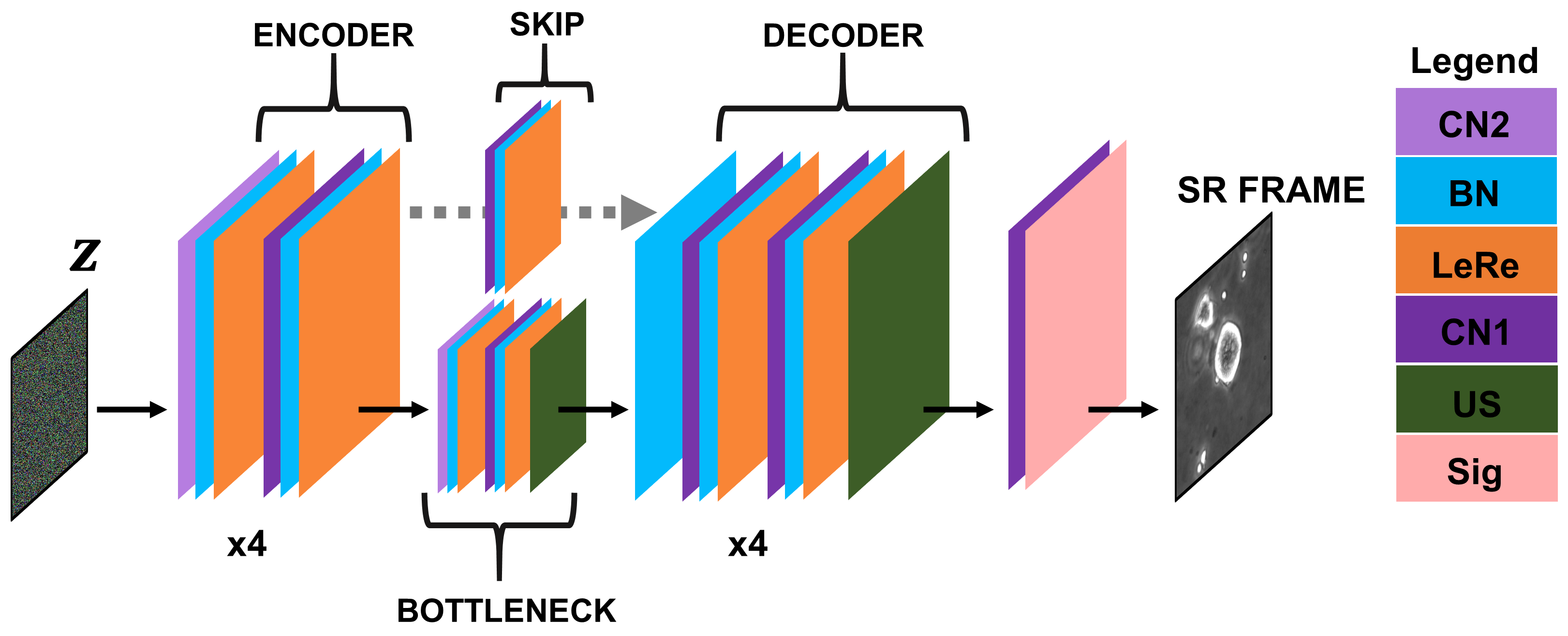}
	\caption{\textbf{The used Encoder-Decoder Skip via concatenation architecture.} In the legend on the right: CN2 stands for Convolutional layer with stride 2, BN for batch normalization, LeRe for Leaky ReLu activation, CN1 for Convolutional layer with stride 1, US for Lanczos Upsampling, Sig for sigmoid activation. The image $z$ is the random input image of the RDPV method and the SR frame is the computed output.}
	\label{fig:net}
\end{figure}

\begin{figure*}
    \centering
	\includegraphics[width=0.9\textwidth]{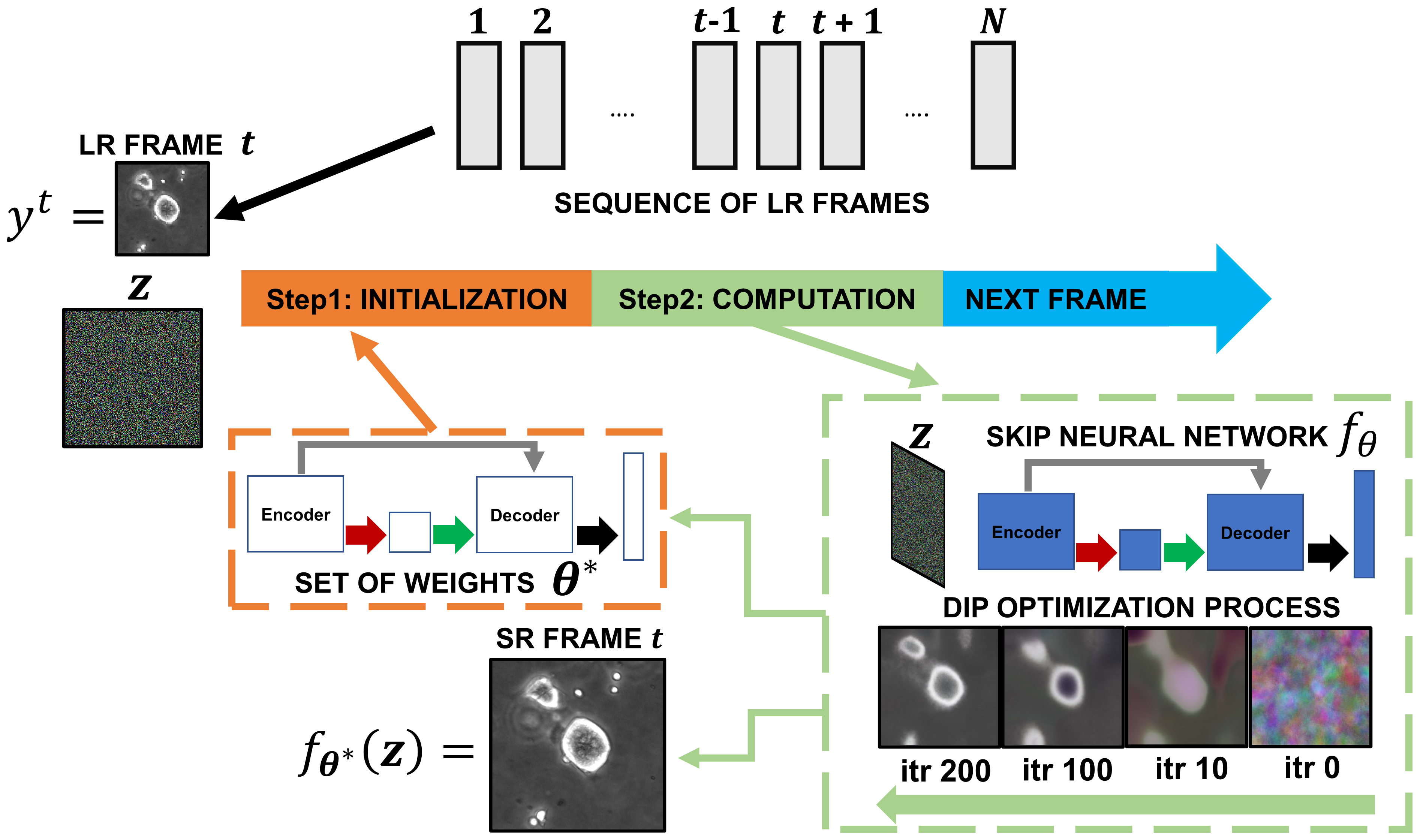}
	\caption{\textbf{RDPV scheme.} The input of the proposed RDPV are the sequence of TLM video frames, the random input image $z$ and the neural network architecture $f_{\theta}$. The method is divided in two steps: the initialization (orange) and the computation (green) step. At time $t$, given the input frame $y^{t}$, the neural network $f_{\theta}$ is initialized by the set of weights which are the output of the computation step at time $t-1$. Then, the classical DIP iterative optimization process is carried out until the considered stopping rule is satisfied. The output of the computation step are the set of weights $\theta^{*}$ which are used for the initialization of $f_{\theta}$ at time $t+1$ and the SR frame at time $t$ computed as $f_{\theta^{*}}(z)$.}
	\label{fig:UDPV_scheme}
\end{figure*}

Motivated by the interesting properties of DIP in the SISR field, we make use of this unsupervised learning-based technique to increase the spatial resolution of TLM videos. The used network is an Encoder-Decoder CNN architecture with long skip connections implemented via concatenation \citep{Ulyanov}. The encoder and decoder parts are made up of four encoder and decoder base units, respectively. The encoder base units accomplish convolutions with 128 features maps, batch-normalization layers and Leaky Relu activations. The downsampling is performed by the convolutional layers setting the stride equals to two.  The decoder base units consider again convolutions with 132 features maps, batch-normalization layers and Leaky Relu activations and in addition an up-sampling Lanczos operator is introduced. The long skip connections by concatenation added between the encoder and the decoder base units make use of convolution with 4 feature maps and batch normalization.  The architecture and the transformations involved are sketched in Fig. \ref{fig:net}. \\ 
We now explain in detail the proposed methods and their main features. All of them apply one frame at a time. The input requirements are the LR frames, a fixed random image $z$ and the CNN architecture properly initialized. We point out that for all the implemented methods, the whole process can be divided into two steps: the initialization and the computation steps. The initialization step defines the starting set of weights initializing the fixed CNN architecture. The computation step involves the iterative process solving a fixed optimization problem by means of standard iterative gradient-based algorithms \citep{bottou2010large,kingma2014adam}. We refer to our first proposal as DPV algorithm. This method treats the frames constituting the TLM videos as uncorrelated images initializing randomly the fixed CNN architecture. The computation step solves the optimization problem \eqref{DIP_minimization} and as stopping criterion a maximum number of iterations is fixed. We remark that DPV can be seen as a simple generalization of the standard DIP to the case of videos. 
Our second proposal is the RDPV method. A scheme of the RDPV is depicted in Fig. \ref{fig:UDPV_scheme}. It differs from the previous DPV algorithm both in the initialization and the computation phase. In order to exploit the temporal correlations among neighboring frames the fixed Encoder-Decoder with skip concatenations architecture applied to the frame at time $t+1$ is initialized by the set of weights $\theta^{*}$ computed during the computation phase related to the frame at time $t$. The computation step, figured in Fig. \ref{fig:UDPV_scheme} (green square), differs from the DPV since the iterative process is early-stopped by means of a criterion which looks at a fixed window of computed values of the objective function in \eqref{DIP_minimization} and ends the iterative process when the values flatten (for more details see paragraph \ref{simulation_settings}). These novelties lead to better results in terms of quality and computational cost saving with respect to the DPV method. Our third proposal is inspired by \citep{liu2019image} where following the classical reconstruction based approaches \eqref{eq:rec_based}, the authors add Total Variation priors \citep{Rudin1992} of the form: 

\begin{equation}   \label{eq:TV_reg}
TV_{p}(u)= \sum_{i=1}^{n} \left( |(D_h u)_i|^p + |(D_v u)_i|^p\right)^{1/p},
\end{equation}

where $u$ is a generic 2D image, $n$ is the number of pixels and $D_{h}$, $D_{v}$ are the first order finite difference discrete operators along the horizontal and vertical axes, respectively.
Hence, our third method attempts to solve in the computation step the optimization problem:  

\begin{equation} \label{DIP_minimization_TV}
    \theta^{*}= \arg \min \lVert y - S f_{\theta}(z) \rVert^{2}_{2} + \lambda TV_p(f_{\theta}(z)) \quad x^{*}=f_{\theta^{*}}(z),
\end{equation}

The parameter $\lambda$ is a positive hand-tuned parameter which represents a trade-off between the quadratic term and the $TV_{p}$ regularizer. Setting $p=1$ or $p=2$ in \eqref{eq:TV_reg}, we obtain the anisotropic and the isotropic TV, respectively. Both are usually employed in SR to suppress noisy components and reconstruct HR images with sharp edges \citep{yue2016}.  The anisotropic regularizer splits the contribution of the gradient components while they are jointly involved in the isotropic one. We stress that in \citep{liu2019image} the authors use the anisotropic TV, while we also consider the isotropic formulation. We underline that this approach exploits the TV regularizers in order to improve the performances obtained by the standard RDPV. We refer to the methods solving \eqref{DIP_minimization_TV} with $p=1$ and $p=2$ as RDPV-TVa and RDPV-TVi, respectively. We remark that RDPV-TVa and RDPV-TVi differ from RDPV only in the computation step. Moreover, we observe that all the proposed methods are unsupervised since we do not need any training set. Finally, we stress that DPV has been considered in order to underline the benefits carried out by the RDPV and its TV-based variants.


\section{Materials and Data Analysis\label{materials}}

\subsection{Synthetic Cell videos}
The dataset of synthetic videos analyzed in the present paper is based on the work by \citep{Comes2019}, where the authors derive a stochastic particle model to artificially mimic the migration of immune cells towards a target tumor cell and their consequent interaction \citep{vacchelli2015chemotherapy}. Immune cells are considered as single entities which migrate according to a random walk with drift $\lvert \mu \rvert$, constant in modulus. However, at the same time, they interact among them and with a target tumor cell. The tumor-immune interaction is modelled as a repulsive-attractive potential \citep{sepulveda2013collective} acting for a priori set time,$T_{eff}$, known as effectiveness time. The values of the drift modulus and the effectiveness time are chosen in accordance with estimations from real experiments \citep{vacchelli2015chemotherapy}. A total amount of 100 synthetic videos are generated using MATLAB  R2017b (MathWorks, Natick, MA). Each video represents a region of interest of $288\times288$ pixels containing 16 immune cells and one target tumor cell. A total number of 100 frames is collected for each video emulating a video frame acquisition of 20 seconds. The theoretical trajectories of the total 1600 immune cells (16 for each of the 100 synthetic videos), obtained as result of the implementation of the stochastic particle model, are here considered as ground truth trajectories (GT trajectories). 
In the following, we will refer to the described atlas of videos as original or noise-free synthetic videos or simply as synthetic videos. 
The original videos are further corrupted by adding white Gaussian noise with standard deviation $\sigma = 0.001$. 
In the following, we will refer to the latter videos as corrupted synthetic videos.

\subsection{Real Cell Videos}
The proposed approach is validated on two sets of real cell videos, namely Videos Type 1 and Videos Type 2, each one containing 10 videos, respectively. Videos Type 1 and Type 2 are sets of real cell videos from experiments described in a paper by \citep{nguyen2018dissecting} where 3D co-cultures in microfluidic devices embedding 4 cell populations (cancer, immune, fibroblasts, and endothelial) recreate ex vivo a human tumor ecosystem (HER2+ breast cancer). For both video sets the drug was Trastuzumab (Herceptin, Roche) at 10 mg/mL concentration. Both video sets include untreated control conditions (5 videos) and drug-treated conditions (5 videos). 
For the Videos Type 1 the two conditions, untreated and drug-treated, are run in parallel, in two different microfluidic devises. The drug is added in the culture medium. Images are taken in parallel using multi-positioning acquisition mode. For the Videos Type 2 the two conditions, untreated and drug-treated, are run sequentially, in the same microfluidic devise. The drug is injected inside the microfluidic device by a syringe pump (1 ml/min flow during 1.5 hrs). Videos are acquired before (untreated condition) and after injection (drug-treated condition).
Each video represents a Region of Interest (ROI) of $288\times288$ pixels with a tumor cell occupying its center. Each video frame has been acquired every two minutes. Videos Type 1 count 31 frame while Videos Type 2 are composed by 57 frames.

\subsection{Cell tracking}
To the aim of this work, we track cells within the live cell videos under study using the proprietary tracking tool \textit{Cell Hunter} \citep{biselli2017organs,parlato20173d,Comes2019}, which basically consists of two main stages. First, a localization/segmentation of the cell candidates using the Circular Hough Transform (CHT) \citep{davies2004machine} allows to estimate cell centers and radii. Among various localization approaches, the CHT has been preferred due to its capability to brilliantly solve the problem of cell occlusion (due to projection errors for example of the 3D scene in the 2D domain) and to properly work in presence of very small objects (immune cells can be represented by a very low amount of pixels). 
Second, cell trajectories are constructed by linking cell nuclei centers detected along the video frames thanks to an enhanced version of Munkres’ algorithm \citep{munkres1957algorithms} for Optimal sub pattern Assignment Problems (OAP). 

\subsection{Statistical Analysis}
A two-fold evaluation process is carried on. Both synthetic and real videos are involved in a test to verify image similarity between original HR and SR video frames based on the computation of two metrics: PSNR and SSIM. For both synthetic and real videos, the average PSNR and the average SSIM values computed over all the video frames are calculated. Since real videos are acquired by inverted microscopes (\textit{see} Section 4.3), they are affected by undesired effects, e.g., additive noisy patterns, related to external factors or intrinsic limitations of the acquisition instrument \citep{di2016time}. Under this consideration, we filter the acquired frames with a smoothing Gaussian kernel and treat them as ground truth \citep{haddad1991class}. 
The PSNR is usually defined in the logarithmic decibel scale (dB) as follows \citep{hore2010}:

\begin{equation}
\mathrm{PSNR} = 20 \log_{10} \left( \dfrac{\max(Y) - \min(Y)}{\mathrm{\sqrt{MSE}}} \right),
\label{eq:param1}
\end{equation}

where MSE is the averaged square difference between the original HR video frame $Y$ and the super resolved video frame $X$. 

The higher the PSNR is, the better the image quality is. 
Anyway, since such metric is sample-dependent and does not allow predicting the perceived image quality, the  perceptual metric SSIM is also used \citep{hore2010}:

\begin{equation}
\mathrm{SSIM}=\dfrac{\left( 2 \mu_{X}\mu_{Y} + c_{1} \right) \left( 2 \sigma_{X}\sigma_{Y} + c_{2} \right)}{\left( \mu_{X}^{2} + \mu_{Y}^{2} + c_{1} \right) \left( \sigma_{X}^{2} + \sigma_{Y}^{2} + c_{1} \right)},
\label{eq:param2}
\end{equation}

where $X$ and $Y$ are the super resolved and the original HR images of size $L M \times L N$ respectively, whilst  $\mu_X, \mu_Y$, $\sigma_X,\sigma_Y$ are the mean and standard deviation of $X$ and $Y$ and  $c_i=(max(Y) k_i), \ k_i \ll 1, \ i=1,2$ are constants. SSIM can take values in the range $[0,1]$. The larger the values is, the greater the visual image quality is. 
The original synthetic videos are also utilized to evaluate how the performance of the tracking software  is affected by the proposed SR algorithm. More specifically, immune cell trajectories detected from the tracking algorithm on LR videos (LR trajectories), on original HR videos (HR trajectories) and on SR videos (SR trajectories) are compared with the theoretical/ground truth trajectories (GT trajectories), i.e., trajectories directly obtained from the implementation of the stochastic particle model in the creation phase of the synthetic videos \citep{Comes2019}. This kind of analysis is accomplished only on synthetic videos because theoretical trajectories are not available in real contexts. The tracking algorithm could fail by missing trajectories or by associating tracts corresponding to different detected trajectories to a single ground truth trajectory. For this reason, tracking performances are assessed by computing the percentage of detected cell trajectories with respect to the total number of the ground truth ones and the so-called swapping error \citep{huth2010significantly} that measures the average number of swaps per trajectory. 
Two parameters are then drawn out from LR, HR, SR and GT immune cell trajectories: the ensemble-averaged Mean Square Displacement (MSD) \citep{ernst2014probing} and the mean interaction time with the tumor cell \citep{parlato20173d}. The descriptors extracted from the ground truth trajectories are compared to those extrapolated from LR, HR and SR trajectories. 
The ensemble-averaged MSD captures the overall diffusive attitude in cell movement and it is expressed as
\begin{equation}
                   MSD_{t}= \langle d(r_{i}^1,r_{i}^t)^2 \rangle                                                                
\end{equation}
where the index $i$ denotes the $i^{th}$ track of length $T_i$, $r_{i}^t$ is the position of the $i^{th}$ track at time $t$, with $t=1,…,max(T_i)$, $d$ stands for the Euclidean distance, $\langle \rangle$ indicates the mean operator applied to the Euclidean distance. The agreement between pairs of time-continuous MSDs, namely, MSD for GT trajectories \textit{versus} MSD for LR/HR/SR trajectories, was measured by the mutual concordance correlation coefficient (CCC) \citep{carrasco2003estimating}, whose values fall in the range $[-1;1]$. The greater the CCC value is, the greater the consensus between MSDs is.
Finally, the mean interaction time (MIT) indicates the average number of frames in which each immune cell remains close to the tumor cell, within the interaction radius, defined as twice the sum of tumor and immune cell radii \citep{parlato20173d,Comes2019}. Pairs of MIT distributions, namely, MIT distribution for GT trajectories \textit{versus} MIT distribution for LR/HR/SR trajectories, are compared using the Student's t test. We assume that a p-value lower than $0.05$ indicates a statistically significant difference between the two distributions under comparison. 

\subsection{Simulation Settings}\label{simulation_settings}

An automatic early stopping procedure is implemented to avoid overfitting. The core idea is to monitor the objective functions in  \eqref{DIP_minimization} for the RDPV, and \eqref{DIP_minimization_TV} for the RDPV-TVa and the RDPV-TVi to decide when to stop the algorithm. This technique creates a window (patience) of consecutive values of the objective function and then, if there is not enough decrease during the iterative process, the algorithms are stopped before the fixed maximum number of iterations. 
For the first image of synthetic videos, we impose a maximum of 1000 iterations with early stopping starting from 500 iterations (patience = 50). For the other images, because of the updating rule of the weights, the number of iterations is reduced: we set a maximum of 500 iteration with early stopping starting from 300 iterations (patience = 50). 
For the first image of real videos, we impose a maximum of 3000 iterations with early stopping starting from 2000 iterations (patience = 50). For the other images, because of the updating rule of the weights, the number of iterations is reduced: we set a maximum of 2000 iterations with early stopping starting from 1000 iterations (patience = 50). For a fair comparison, we set the maximum number of iterations for DPV as the number of iterations performed by RDPV according to the aforementioned stopping criterion.
All the videos have been downsampled by a factor equal to $4$ and then reconstructed by SR methods. Therefore, the magnification factor $L$ is imposed equal to 4.  Conversely to other lower factors, this choice allows to effectively reduce the spatial burden due to the storage of TLM videos.  Moreover, we compare the results achieved by our method using a magnification factor for which learning-based methods have shown outstanding results. 
All the simulations are performed on a PC Intel(R) Core(TM) i7-8700U CPU $@$ 3.70 GHz 3.70GHz with 32Gb RAM using Python (v3.6.5) and MATLAB R2018b (MathWorks, Natick, MA) on two NVIDIA GeForce GTX 1080 Ti (11 Gb of memory per GPU) boards with CUDA driver (v9.0) and cuDNN library (v7.1.4).

\section{Results and discussion}\label{results} 

\subsection{The importance of super resolution on cell tracking and descriptors extraction}
\begin{figure}[!t]
\centering
\includegraphics[scale=0.7]{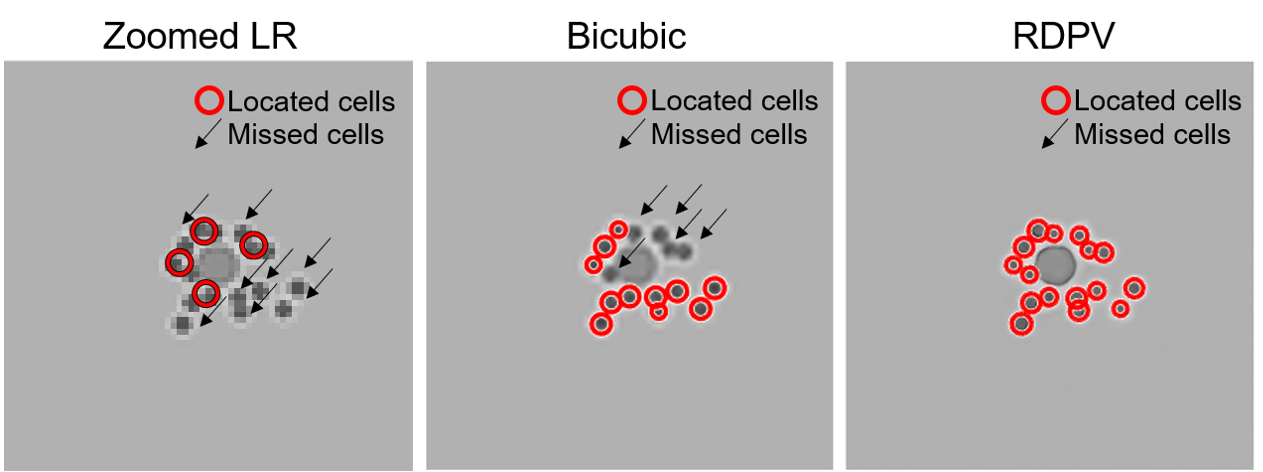}  
\caption{\textbf{Example of cell localization by means Cell Hunter software on a LR frame (left panel) and the corresponding SR counterparts reconstructed by Bicubic (central panel) and the proposed RDPV method (right panel).}}
\label{fig:localizzazione}
\end{figure}

Video analysis usually starts with cell localization and tracking (\textit{see} Section 4.3). The errors made by tracking algorithms in this phase can propagate and heavily compromise the extrapolated biological conclusions. As demonstrated elsewhere \citep{Comes2019}, when the spatial resolution is low, the tracking software could fail in its localization task. In order to corroborate this, in Fig. \ref{fig:localizzazione} we show how the localization algorithm (CHT) acts on LR and SR frames obtained by a coarse SR algorithm and the proposed RDPV, respectively. More precisely, the left panel of Fig. \ref{fig:localizzazione}  depicts an example of cell localization on the LR synthetic video frame where localized cells are marked as red circles and missed cells are pointed out by black arrows. As a result, we observe that some cells are missed, and others are identified as unique entities because partially overlapped. The most simple and popular choice for image up-sampling is the bicubic interpolation algorithm \citep{keys1981cubic}, that is frequently implemented in data analysis software since it quickly provides up-scaled images. In the case of videos, such algorithm is usually applied frame by frame. Despite its rapidity and low computational cost, the bicubic outcome reveals to be blurred with the contour of the cell not sharp. As a consequence, this smoothing effect damages the edge detection of the circular-shaped objects by the CHT. Indeed, as depicted in the central panel of Fig.\ref{fig:localizzazione}, the lack of fine details may lead to inaccurate localization. It is evident how the software is able to distinguish cells but misses some of them.  As shown in the right panel of Fig. \ref{fig:localizzazione}, a more sophisticated SR algorithm such as the proposed RDPV, allows increasing the trustworthiness of tracking software in localizing cells so overcoming the limitation of coarse results obtained by the standard bicubic algorithm. More precisely, the localization algorithm (CHT) provides better performance on higher resolution images because it is based on the concept of the accumulation point \citep{davies2004machine}. 

\begin{figure}[!t]
\centering
\includegraphics[scale=0.45]{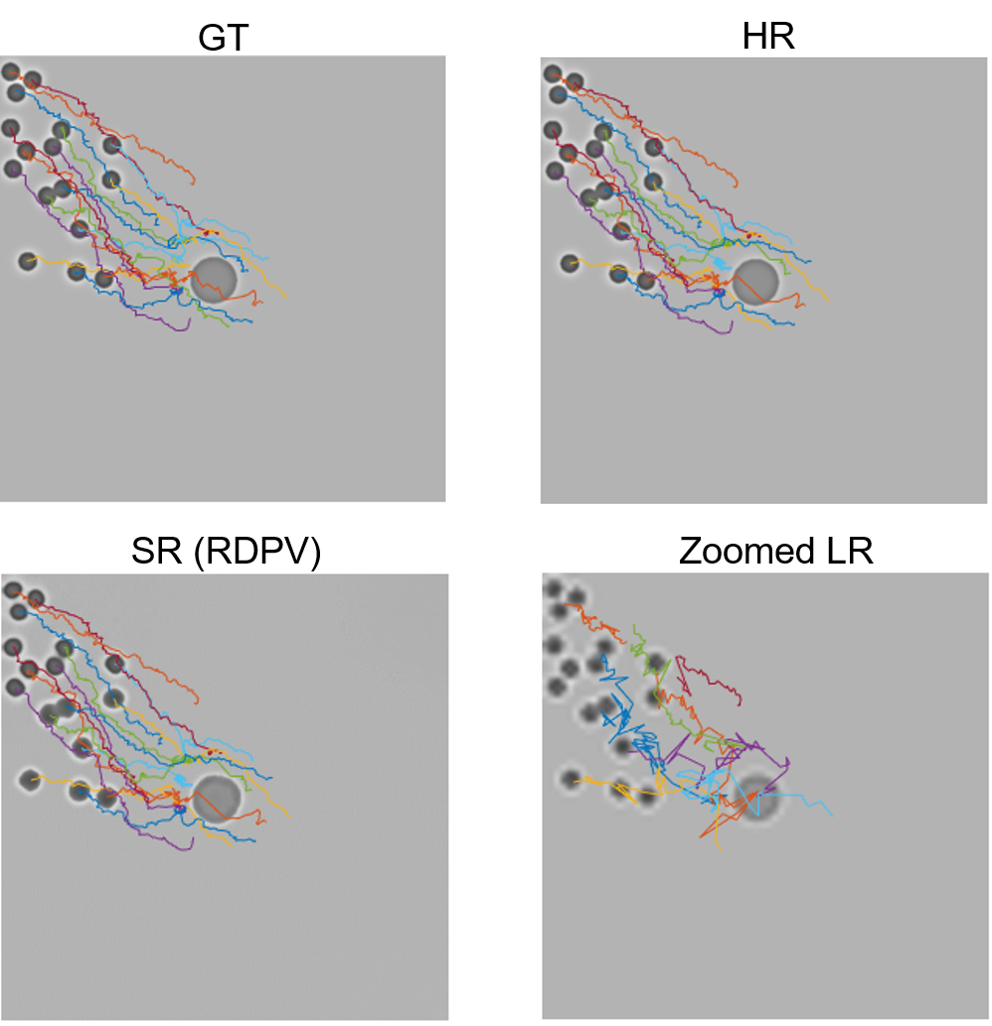}  
\caption{\textbf{Visual representation of immune cell trajectories on a synthetic video.} Ground-truth (GT) trajectories directly obtained from the implementation of the model in the creation phase of the synthetic videos(upper-left panel). High resolution (HR) trajectories identified by Cell Hunter software on the original HR video (upper-right panel). Super resolution (SR) trajectories identified by Cell Hunter software on the SR video, reconstructed by the proposed RDPV method  (lower-left panel). Lower resolution (LR) trajectories identified by Cell Hunter software on the LR video (lower-right panel). A zoom of LR trajectories is provided for a better visualization (Zoomed LR).}
\label{Fig:tracce}
\end{figure}

\begin{figure}[!t]
\centering
\includegraphics[scale=0.7]{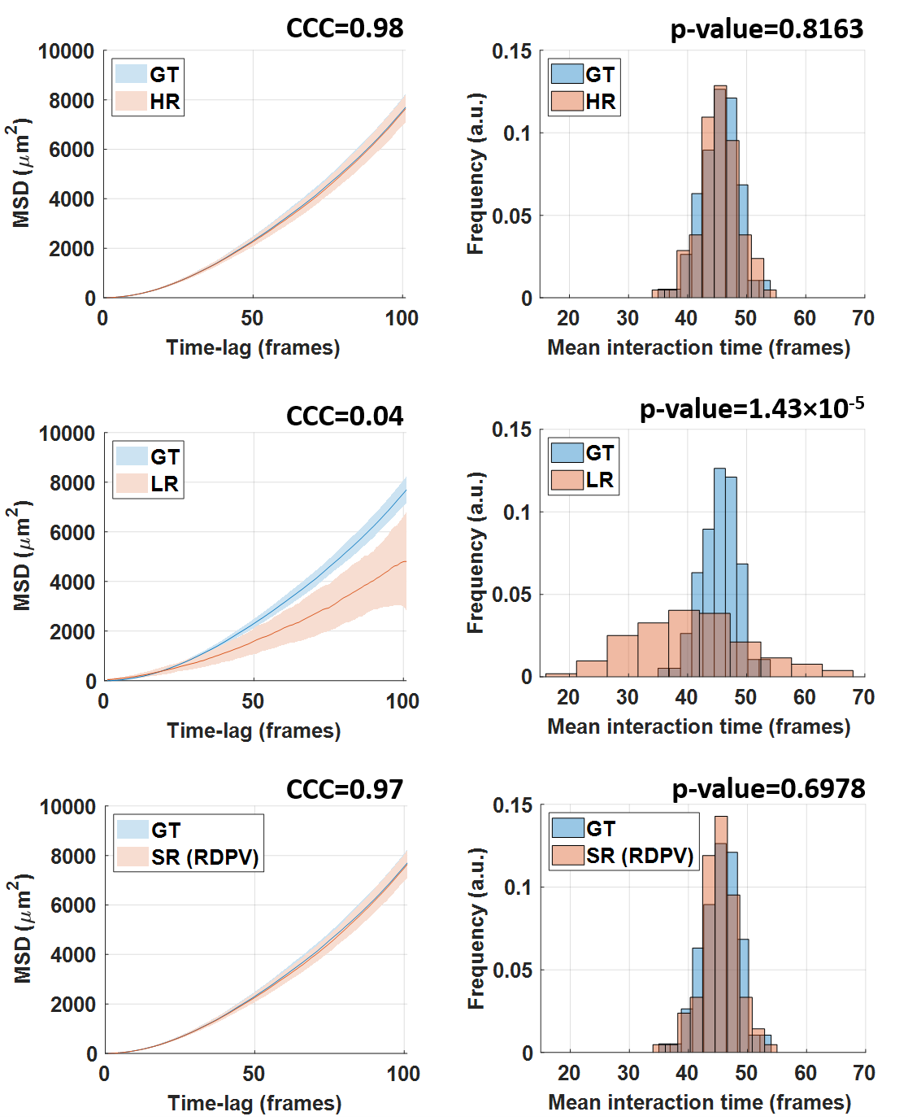}
\caption{\textbf{Performance evaluation in terms of motility and interaction descriptors on synthetic videos.} (left panel) Comparison between couple of Ensemble-Averaged MSD curves: GT vs HR (upper-left), GT vs LR (centre-left ), GT vs SR (lower-left).CCC values are also highlighted. (right panel) Comparison between couple of Mean Interaction Time distributions: GT vs HR (upper-right), GT vs LR (centre-right ), GT vs SR (lower-right). P-values for the Student's t-test are also indicated. In the legend GT stands for Ground Truth, HR stands for High Resolution, SR stands for Super Resolution, LR stands for Lower Resolution.}
\label{Fig:distribuzioni}
\end{figure}

It is interesting to investigate how the performance of the tracking software and the consequent extraction of some motility and interacting descriptors are affected by super resolved video frames. 
A visual comparison between GT, HR, SR and, LR trajectories on synthetic video frames is provided in Fig. \ref{Fig:tracce}. A zoom is supplied for a better visualization of the LR trajectories. Conversely to LR trajectories, the appearance of HR and SR trajectories is very similar to that of GT ones. From a quantitative point of view, the percentage of total number of detected LR trajectories on the 100 synthetic videos with respect to the overall number of the GT ones (1600) is equal to $51 \%$, while it reaches the $100 \%$ for the detection of both the HR and the SR trajectories. The swap error counts  $18.4$, $3$ and $3.5$ swaps per trajectory, for the LR, HR and SR trajectories, respectively. This is a consequence of the previously underlined CHT errors made on LR video frames (Fig. \ref{fig:localizzazione}) since the estimation of trajectories is strictly correlated to the detection of cells frame by frame. 
In order to prove that the performance of the tracking software reflects on the consequent feature extraction, we then compare the motility and interaction descriptors (average-ensemble MSD and Mean Interaction Time, MIT) extracted from GT trajectories with those calculated from HR, SR and LR trajectories, respectively. In left panels of Fig. \ref{Fig:distribuzioni}, pairs of average-ensemble MSD curves are drawn: the GT curve is juxtaposed on the HR curve in the upper panel, on the LR curve in the central panel, and on the SR curve in the lower panel. The GT and HR curves as well as GT and SR curves are almost completely overlapped producing high CCC values. The LR curve, instead, significantly differs from the GT one resulting in a low CCC value. All the CCC values are highlighted on the respective left panels in Fig. \ref{Fig:distribuzioni}. Furthermore, in the right panel of Fig. \ref{Fig:distribuzioni}, couple of MIT distributions in comparison are shown: GT and HR distributions in the upper panel, GT and LR distributions in the central panel, and GT and SR distributions in the lower panel. The p-value for the Student's t-test is computed for all the three comparative scenarios. The information carried by HR and SR distributions is reliable (p-values $>$ 0.05), whereas the one related to LR videos is completely biased. The flattening of the LR distribution is an effect of the low spatial resolution: cells are hardly localized and hence the effective number of interacting cells decrease. In this latter case, distributions appear significantly different (p-value $<$ 0.05), despite they represent the same experimental condition.
This test confirms that analyzing videos reconstructed by the proposed SR algorithm leads to a minimum information loss with respect to analyzing HR videos and at the same time a great information gain with respect to analyze LR videos. 

%
%
%
%
\subsection{Image quality evaluation on synthetic cell videos }
\begin{figure}[h]
\centering
\includegraphics[width=0.4\textwidth]{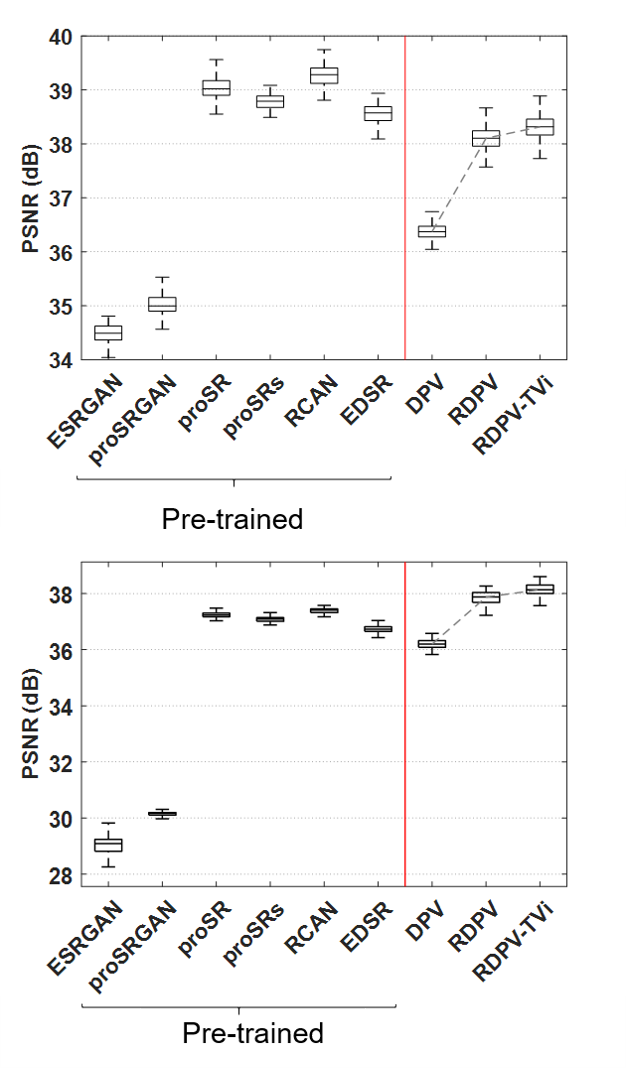}  
\caption{\textbf{Quantitative results in terms of PSNR on synthetic videos ($\sigma = 0  $ , bottom panel) and their corrupted counterparts ($\sigma = 0.001  $, lower panel).} Boxplots comprise the average values of PSNR computed over all the frames of each synthetic video.}
\label{fig:PSNR_artificiali}
\end{figure}

So far, we have underlined the importance of the high spatial resolution for TLM videos in order to get a successful tracking and extraction of kinematic and interaction descriptors. In Fig. \ref{fig:localizzazione}, we test the CHT on the SR output of the bicubic algorithm, highlighting that sophisticated SR algorithms are required in order to retrieve fine image details.   
As exposed in Section 2, some state-of-the-art methods address the problem of SR. To fulfill the task, most of them are based on neural networks and require training on a given dataset. We compare video restoration provided by some of the best perfoming trained methods, ESRGAN \citep{ESRGAN}, proSRGAN, proSR, proSRs \citep{proSR}, RCAN \citep{RCAN}, EDSR \citep{EDSR}, with those obtained by the proposed non regularized (DPV and RDPV) and regularized (RDPV-TVi) unsupervised learning algorithms. Performances are evaluated on either original or corrupted synthetic videos (\textit{see} Section 4.1). Figure \ref{fig:PSNR_artificiali} shows distributions of the average PSNR values computed over all the frames of each of the 100 synthetic videos for all the above mentioned methods. The top panel of Fig. \ref{fig:PSNR_artificiali} refers to original synthetic videos. The bottom panel of Fig. \ref{fig:PSNR_artificiali} refers to corrupted synthetic videos. The average SSIM values with the relative standard deviations are reported in Table \ref{Table:SSIM_artificiali}: the first column is related to original synthetic videos; the second column is related to the corrupted synthetic videos. The proposed RDPV and RDPV-TVi algorithms always outperform the standard DPV. This is because RDPV implements the new recursive updating rule which takes into account the knowledge acquired by the previous reconstructed frames. Moreover, the regularization term within RDPV-TVi adds to the model additional a priori information on the solution which is not completely captured by the fixed CNN architecture. In such a way, we attenuate one of the main flaws of the standard DIP framework, that is the choice of the network architecture. 
Very promising results are achieved from the comparisons with the trained networks. Comparable performances are observed on noise-free artificial videos (bottom panel of Fig. \ref{fig:PSNR_artificiali}). Better performances are achieved on corrupted artificial videos (lower panel of Fig. \ref{fig:PSNR_artificiali}). This is because one of the main drawbacks of trained architecture is the instability with respect to the presence of noise components in the incoming data \citep{wang2020}. They are not able to filter out the added noise from test images if the training dataset does not present a considerable number of noisy-images at different levels of noise. Anyway, this necessity might be limiting in practical applications. The presented results stress again the importance of developing an algorithm whose output is not dependent on a fixed set of image examples. According to such quantitative results, we depict images obtained by the worst and the best trained methods (ESRGAN and RCAN, respectively) alongside GT,  LR, DPV, RDPV and RDPV-TVi images. 
More specifically, Fig. \ref{fig:sintetic_reconstruction} shows a noise-free synthetic video frame while Fig. \ref{fig:noisy_reconstructions} highlights one of the video frames for the corrupted synthetic videos. For the noise-free synthetic videos, RCAN achieves the best performance in terms of average PSNR and SSIM but, as emerged from Fig.\ref{fig:sintetic_reconstruction}, the immune cells shape appears distort (red and green arrows). The same warping also holds for the ESRGAN method. The standard DPV, instead, is not able to guarantee a fine result, conversely to RDPV which, on its part, has the drawback to not effectively separate some of the immune cells (red and green arrows). The addition of TV regularization (RDPV-TVi) leads to optimal results both in terms of immune cell shape and differentiation. For the corrupted synthetic videos in Fig. \ref{fig:noisy_reconstructions}, these remarks are still valid.

\begin{table}[!h]

\begin{center}
\caption{\textbf{Quantitative results in terms of SSIM average values on synthetic videos ($\sigma = 0  $ ) and their corrupted counterparts ($\sigma = 0.001  $).}  The average values of SSIM and the corresponding standard deviations are highlighted. } 

\scalebox{1}{
\begin{tabular}{c|c|c}\hline 
                   \multicolumn{3}{c}{\textbf{Average SSIM}} \\
\hline
\textbf{Methods}    &  $\sigma = 0  $               & $\sigma = 0.001  $   \\
\hline 
ESRGAN     & 0.9592 $\pm$ 2.2 $\times$ $10^{-3}$ & 0.5781 $\pm$ 1.7 $\times$ $10^{-2}$   \\ \hline
proSRGAN  & 0.9833 $\pm$ 7.7 $\times$ $10^{-4}$ & 0.6742 $\pm$ 5.0 $\times$ $10^{-3}$  \\ \hline
proSR      & 0.9916 $\pm$ 5.3 $\times$ $10^{-4}$ & 0.9553 $\pm$ 1.4 $\times$ $10^{-4}$  \\ \hline
proSRs     & 0.9924 $\pm$ 3.7 $\times$ $10^{-4}$ & 0.9532 $\pm$ 1.7 $\times$ $10^{-4}$  \\ \hline
RCAN       & 0.9924 $\pm$ 4.0 $\times$ $10^{-4}$ & 0.9546 $\pm$ 1.7 $\times$ $10^{-4}$  \\ \hline
EDSR       & 0.9913 $\pm$ 1.3 $\times$ $10^{-4}$ & 0.9549 $\pm$ 1.9 $\times$ $10^{-4}$  \\ \hline
DPV        & 0.9806 $\pm$ 8.9 $\times$ $10^{-4}$ & 0.9765 $\pm$ 1.1 $\times$ $10^{-3}$  \\ \hline
RDPV       & 0.9906 $\pm$ 7.3 $\times$ $10^{-4}$ & 0.9886 $\pm$ 1.0 $\times$ $10^{-3}$  \\ \hline
RDPV-TVi   & 0.9920 $\pm$ 6.5 $\times$ $10^{-4}$ & 0.9916 $\pm$ 4.2 $\times$ $10^{-4}$ \\ \hline

\end{tabular}\label{Table:SSIM_artificiali}
}
\end{center}

\end{table}

\input{img_tex/fig3_2}
\input{img_tex/fig4}

\subsection{Image quality evaluation on real cell videos}

\begin{figure}[!t]
\centering
\includegraphics[width=0.4\textwidth]{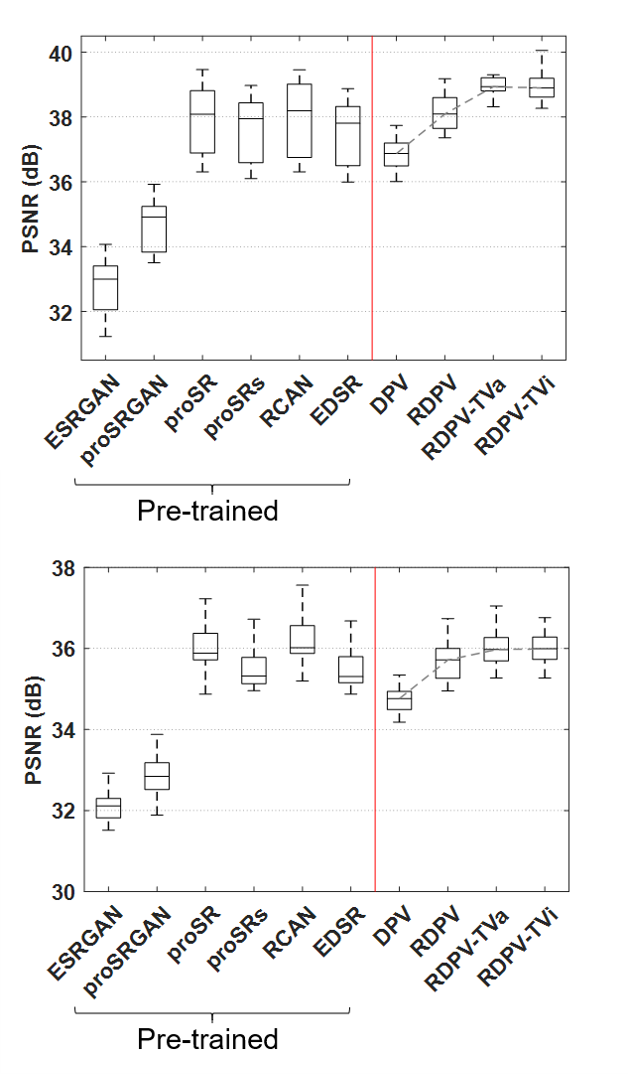}  
\caption{\textbf{Quantitative results in terms of PSNR on real videos Type1  (bottom panel) and Type2 (lower panel).} Boxplots comprise the average values of PSNR computed over all the frames of each real video. }
\label{fig:PSNR_video_reali}
\end{figure}

We finally validate the proposed approach on videos from two diverse OOC experiments by exploiting tumor-immune interaction: Videos Type1 and Videos Type2. For details about videos, please refer to Section 4.2. As for the synthetic videos, we compare the quality of ground truth frames with those achieved by the proposed methods, without and with regularization, i.e., DPV, RDPV and RDPV-TVi/RDPV-TVa, respectively, also including the trained methods. Figure \ref{fig:PSNR_video_reali}  and  Table \ref{tab:SSIM_video_reali} show the average PSNR values and average SSIM values with the relative standard deviations for both Video Type 1 and Video Type  2, respectively. The average PSNR values are computed over all the frames of each of the real videos of the two types. The average values of SSIM are calculated over all the videos. 
The proposed methods RDPV,RDPV-TVi and RDPV-TVa outperform the standard DPV both for Video Type 1 and Video Type 2 in terms of average PSNR and SSIM. We stress that the addition of both isotropic and anisotropic Total Variation improve the performances of the RDPV, confirming once again the importance of these additional terms. For what concerns the comparisons with the trained architecture,  RDPV, RDPV-TVi and RDPV-TVa outperform them in terms of average PSNR on Video Type 1 (top panel of \ref{fig:PSNR_video_reali}) and reach comparable performances on Video Type 2 (bottom panel of \ref{fig:PSNR_video_reali}). According to the SSIM metric our methods reach similar performances with respect to the trained algorithms (Table \ref{tab:SSIM_video_reali}).  

No less relevant, from the PSNR distributions shown in the boxplots in Fig. \ref{fig:PSNR_video_reali}, it is remarkable to see that the distributions of the average PSNR values have a high variance whereas unsupervised methods have more stationary performances over all the tests executed on real videos. This is more evident for the average PSNR distributions on Video Type 1 (top panel of \ref{fig:PSNR_video_reali}). This highlights a stronger sensitivity to the given input frames for the trained methods with respect to the unsupervised ones. We stress that, for each of the two video types, even if the frames recorded belong to the same experiment, some conditions may change from one video or even frames to another such as the brightness of the field of view. Indeed, analyzing the acquired videos, we observe that sudden changes of luminosity are evident especially for Videos Type 1,thus negatively affecting the performances of trained methods in some cases. This confirm the need of unsupervised methods for TLM videos super resolution, since it not feasible to collect a dataset that takes into account all the possible real boundary conditions.
As qualitative evaluation, in Figure \ref{fig:real_video} we depict a video frame example (from Videos Type 2) obtained by the worst and the best trained methods (ESRGAN and RCAN, respectively) alongside GT, LR, DPV, RDPV and RDPV-TVi images. For a better visualization of the edges of the portrayed cells, Fig. \ref{fig:real_video_sobel} shows the edge map of the same video frame example after applying Canny Filter \citep{canny1986computational}. In both Fig. \ref{fig:real_video} and \ref{fig:real_video_sobel}, three regions of interest are closed-up and highlighted by using three different coloured squares: red, green and blue. All the approaches increase the resolution of the starting LR frame and in particular all of them are able to separate the cells belonging to the cluster within the red square. However, as it is evident from the edge map (Fig. \ref{fig:real_video_sobel}, red square), the trained approaches introduce artifacts alongside the cell cluster. Such artifacts can dramatically affect the cell tracking software in the localization phase thus identifying false positive cell candidates. We observe that all the unsupervised methods reach quite similar results with respect to the ground-truth image. However, the cells within the cluster (Fig. \ref{fig:real_video_sobel}, red square) look better separated for the image obtained by RDPV-TVi, thus confirming that the addition of the Total Variation regularizer improves the results. 
\begin{table}[!t]

\begin{center}
\caption{\textbf{Quantitative results in terms of SSIM average values on real videos Type 1 and on real videos Type 2.} The average values of SSIM calculated over all the videos and relative standard deviations are highlighted.}

\scalebox{1}{
\begin{tabular}{c|c|c}\hline 
                   \multicolumn{3}{c}{\textbf{Average SSIM}} \\
\hline
 \textbf{Method}          & Videos Type1 & Videos Type2 \\
\hline
ESRGAN       & 0.8593  $\pm$ 2.1 $\times$ $10^{-2}$  &  0.8775 $\pm$ 9.4 $\times$ $10^{-3}$       \\ \hline
proSRGAN    & 0.9312  $\pm$ 1.0 $\times$ $10^{-2}$  &  0.9243 $\pm$ 9.8 $\times$ $10^{-3}$       \\ \hline
proSR        & 0.9750  $\pm$ 2.7 $\times$ $10^{-3}$  &  0.9778 $\pm$ 1.5 $\times$ $10^{-3}$       \\ \hline
proSRs       & 0.9736  $\pm$ 2.8 $\times$ $10^{-3}$  &  0.9759 $\pm$ 1.5 $\times$ $10^{-3}$       \\ \hline
RCAN         & 0.9759  $\pm$ 2.5 $\times$ $10^{-3}$  &  0.9783 $\pm$ 1.4 $\times$ $10^{-3}$       \\ \hline
EDSR         & 0.9735  $\pm$ 2.9 $\times$ $10^{-3}$  &  0.9767 $\pm$ 1.4 $\times$ $10^{-3}$       \\ \hline
DPV          & 0.9463  $\pm$ 5.0 $\times$ $10^{-3}$  &  0.9356 $\pm$ 7.5 $\times$ $10^{-3}$       \\ \hline
RDPV         & 0.9618  $\pm$ 3.6 $\times$ $10^{-3}$  &  0.9578 $\pm$ 4.2 $\times$ $10^{-3}$       \\ \hline
RDPV-TVa     & 0.9704  $\pm$ 2.8 $\times$ $10^{-3}$  &  0.9673 $\pm$ 2.3 $\times$ $10^{-3}$       \\ \hline
RDPV-TVi     & 0.9702  $\pm$ 2.6 $\times$ $10^{-3}$  &  0.9669 $\pm$ 2.8 $\times$ $10^{-3}$       \\ \hline

\end{tabular} \label{tab:SSIM_video_reali}
}
\end{center}

\end{table}

\subsection{Cell Tracking on SR real videos}
\begin{figure}[h]
\centering
\includegraphics[width=0.6\textwidth]{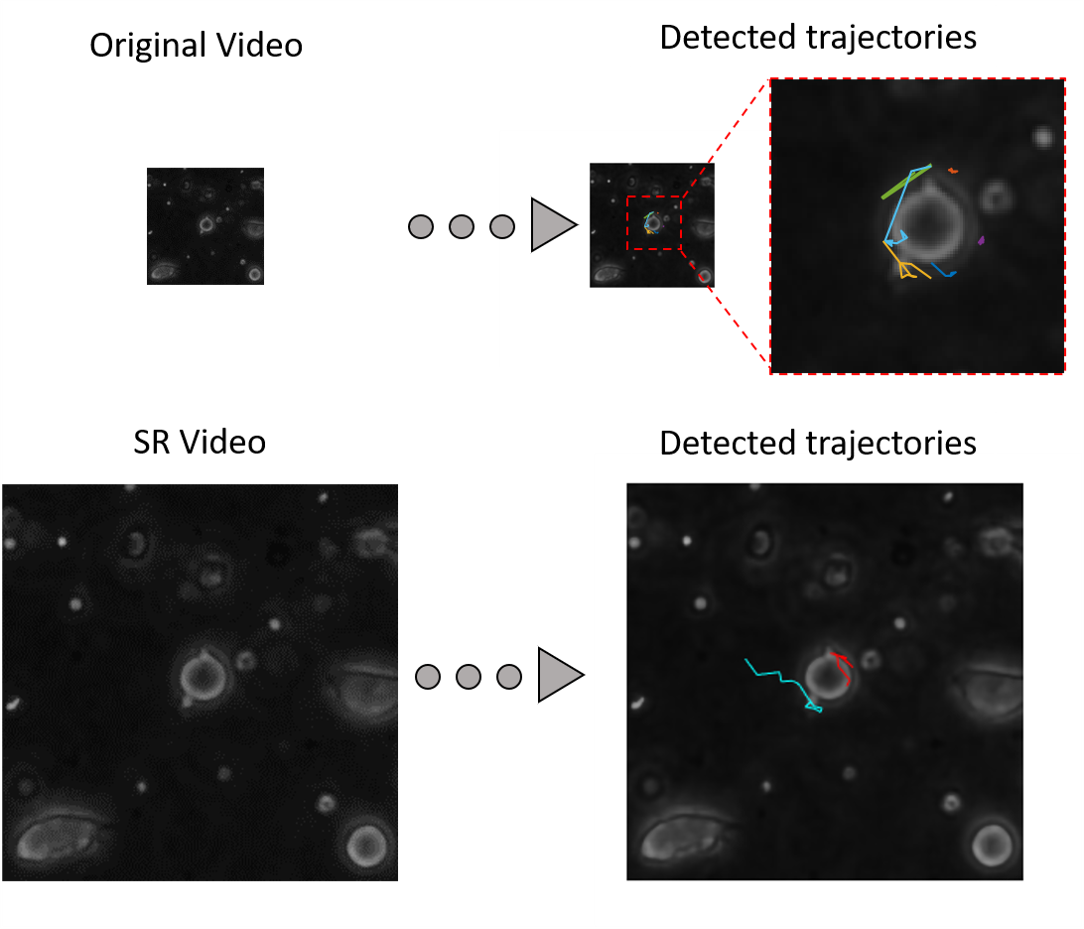}  
\caption{\textbf{Visual representation of detected trajectories by Cell Hunter software on the original LR video and on the SR video reconstructed by the proposed RDPV-TVi method.}}
\label{fig:tracce_reali}
\end{figure}

We have previously carried on a detailed analysis on synthetic videos showing the importance and the efficiency of our method used as preprocessing step on their low resolution counterparts before  tracking and feature extraction are performed. We now test the tracking software on the SR Video Type 1, obtained using the RDPV-TVi method, and on its LR counterpart. Unfortunately, in this case the ground truth trajectories are not available since this is a real video and not a simulated one. In Fig. \ref{fig:tracce_reali}, we compare the LR trajectories with the SR trajectories. The frame is a part of a real video in which two immune cells interact with a tumor cell. In the right upper corner of Fig. \ref{fig:tracce_reali}, the  trajectories detected on the starting low resolution video are represented. As we can observe, the tracking software assigns more than two trajectories to the immune cells for a total of 7 trajectories, which appear, moreover, very segmented and unrealistic. Conversely, giving to the tracking software the SR video computed using the RDPV-TVi method, only two trajectories are detected that results more realistic.


\input{img_tex/fig11}
\input{img_tex/fig12}

\subsection{Comparison with a learning method trained on a dataset of cell images}

\begin{figure}[!t]
\centering
\includegraphics[width=0.4\textwidth]{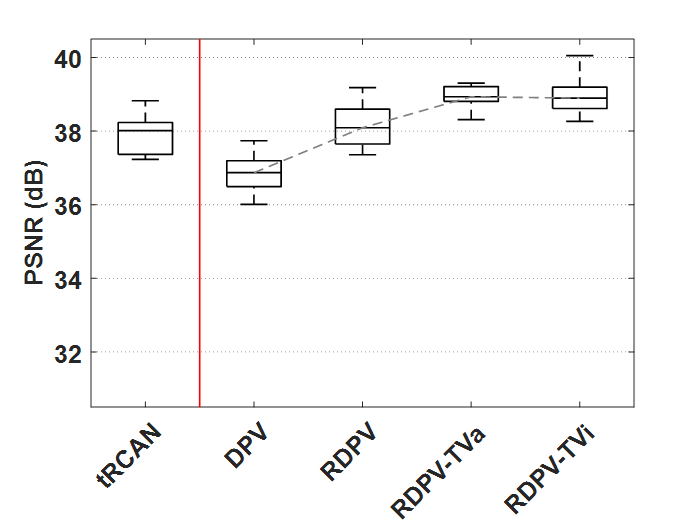} 
\caption{\textbf{Quantitative results in terms of PSNR on videos Type1.} Boxplots comprise  the average values of PSNR computed over all the frames of the 10 videos Type 1. tRCAN indicates RCAN trained on Videos Type2 and some synthetic videos.}
\label{fig:trained_RCAN}
\end{figure}

To conclude the discussion related to real videos, we want to evaluate how the performance of the most performing trained network, RCAN, varies with respect to the training set employed. More specifically, since RCAN weights were taken as a pre-trained model available online \footnote{https://paperswithcode.com/sota/image-super-resolution-on-set14-4x-upscaling}, the question is whether training RCAN on cell images and on images with similar patterns leads to an improvement of the performances. To adequately address the question, by following the same protocol expressed in \citep{RCAN}, we train the RCAN architecture on a new dataset composed by some frames of the synthetic videos and Videos Type2. We refer to this network as tRCAN. Thereafter, we test the tRCAN on Videos Type1 and compare its performance with those of the DPV, RDPV, RDPV-TVi and RDPV-TVa methods. In Fig. \ref{fig:trained_RCAN}, we report the comparison in terms of average PSNR computed on each of the 10 videos. As we can observe from the boxplot in Fig. \ref{fig:trained_RCAN} and by comparing it with the RCAN outcome in Fig. \ref{fig:PSNR_video_reali} (top panel), the usage of a training set made up of cell images reduces the variance of the average PSNR distribution, even if it does not significantly affect the RCAN performance in terms of average PSNR. This confirm how the output of the supervised method RCAN is strongly related to the given dataset. In order to increase significantly the performances of the RCAN, we need to add more cell frames belonging to the Video Type 1, thus resulting not feasible for real applications. As it is evident by the average PSNR boxplot reported in Fig. \ref{fig:trained_RCAN}, the proposed methods RDPV-TVi and RDPV-TVa outperform the tRCAN, and all the other unsupervised methods DPV and RDPV. 

\section{Conclusions}
In this work, we present a novel approach called Recursive Deep Prior Video to overcome the limitation of low resolution in time-lapse microscopy scenario for organ-on-chip applications, within the so-called super-resolution context. The main novelties of the approach refer to the recursive initialization of the weights of the DIP network architecture combined with an efficient early stopping criterion. In addition, the DIP loss function has been penalized by two different Total Variation (TV) based terms. The method has been validated on synthetic, i.e., artificially generated, as well as real videos from OOC experiments related to tumor-immune interaction and compared to the most effective state of the art approaches in the context of trained methods and compared to the proposed extension of DIP for videos (DPV). The proposed approach demonstrates to be feasible to real-time applications due to the unsupervised architecture, robust to noise thanks to the regularized terms, and able to effectively work in combination with state-of-the-art edge localization and edge detection methods for the task of object recognition and biological experiment characterization. Future works will address the problem of improving the effectiveness of the approach in terms of parallelization and of implementation in the routine use of microfluidic devices to accelerate the uptake of OOC experiments.

\bibliographystyle{unsrtnat}






\end{document}